%% file: alignment_paper.tex
\documentclass[sigconf]{acmart}
\setcopyright{none}
\renewcommand\footnotetextcopyrightpermission[1]{}
\usepackage{graphicx}
\usepackage{subcaption}
\usepackage{hyperref}
\usepackage{url}
\usepackage{algorithm}
\usepackage{algorithmic}

\usepackage{amsmath, amsfonts}
\usepackage{bbold}

\newcommand{\RR}{\mathbb{R}}

\newcommand{\calA}{\mathcal{A}}
\newcommand{\calE}{\mathcal{E}}
\newcommand{\calF}{\mathcal{F}}

\newcommand{\bb}[1]{\left( #1\right)}

\newcommand{\bs}[1]{\left\{ #1\right\}}

\newcommand{\norm}[1]{\left\| #1 \right\|}

\usepackage[textsize=tiny,textwidth=1.4cm]{todonotes}

\AtBeginDocument{%
  }

\setcopyright{acmlicensed}
\copyrightyear{2025}
\acmYear{2025}
\acmDOI{}
\acmISBN{}

\begin{document}

\title{Aligning Multilingual News for Stock Return Prediction}

\author{Yuntao Wu}
\email{winstonyt.wu@mail.utoronto.ca}
\orcid{0009-0006-6269-1603}
\affiliation{%
  \institution{University of Toronto}
  \city{Toronto}
  \state{Ontario}
  \country{Canada}
}
\author{Lynn Tao}
\email{lynnyl.tao@mail.utoronto.ca}
\orcid{0009-0000-3856-2146}
\affiliation{%
  \institution{University of Toronto}
  \city{Toronto}
  \state{Ontario}
  \country{Canada}
}
\author{Ing-Haw Cheng}
\email{inghaw.cheng@rotman.utoronto.ca}
\orcid{0000-0002-4872-7888}
\affiliation{%
  \institution{University of Toronto}
  \city{Toronto}
  \state{Ontario}
  \country{Canada}
}
\author{Charles Martineau}
\email{charles.martineau@rotman.utoronto.ca}
\orcid{0000-0002-6896-184X}
\affiliation{%
  \institution{University of Toronto}
  \city{Toronto}
  \state{Ontario}
  \country{Canada}
}
\author{Yoshio Nozawa}
\email{yoshio.nozawa@rotman.utoronto.ca}
\orcid{0000-0002-6395-2688}
\affiliation{%
  \institution{University of Toronto}
  \city{Toronto}
  \state{Ontario}
  \country{Canada}
}
\author{John Hull}
\email{john.hull@rotman.utoronto.ca}
\orcid{0000-0003-4290-3374}
\affiliation{%
  \institution{University of Toronto}
  \city{Toronto}
  \state{Ontario}
  \country{Canada}
}
\author{Andreas Veneris}
\email{veneris@eecg.toronto.edu}
\orcid{0000-0002-6309-8821}
\affiliation{%
  \institution{University of Toronto}
  \city{Toronto}
  \state{Ontario}
  \country{Canada}
}

\renewcommand{\shortauthors}{Y. Wu, L. Tao, I. Cheng, C. Martineau, Y. Nozawa, J. Hull, A. Veneris}

\begin{abstract}
News spreads rapidly across languages and regions, but translations may lose subtle nuances. We propose a method to align sentences in multilingual news articles using optimal transport, identifying semantically similar content across languages. We apply this method to align more than 140,000 pairs of Bloomberg English and Japanese news articles covering around 3500 stocks in Tokyo exchange over 2012-2024. Aligned sentences are sparser, more interpretable, and exhibit higher semantic similarity. Return scores constructed from aligned sentences show stronger correlations with realized stock returns, and long-short trading strategies based on these alignments achieve 10\% higher Sharpe ratios than analyzing the full text sample.
\end{abstract}

\ccsdesc[500]{Computing methodologies~Natural language processing}
\ccsdesc[500]{Applied computing~Economics}

\keywords{international markets, natural language processing, multilingual analysis, optimal transport, return predictions}


\maketitle

\section{Introduction}

Financial markets increasingly reflect a complex interplay of global information flows. News about firms, policy decisions, or macro conditions often appears first in local languages and subsequently in global reporting. 
Translating this multilingual content into English or a single canonical language may obscure subtle inflections, domain-specific nuances, or phrasing differences that carry predictive value. 
When multilingual sources independently cover similar topics, they may emphasize different angles or priorities, so treating them as direct translations risks losing information. 
Finding sementic alignment across different languages could better capture shared meaning, and potentially uncover predictive structure.

Recent advances in natural language processing (NLP) of transformer based large language models (LLMs), especially bidirectional encoder representations from transformers (BERT), suggest a path forward \cite{transformers, bert}. 
\citet{ckx} demonstrate that embeddings from LLMs can extract nuanced, contextual features from news text that outperform traditional bag-of-words and sentiment models in return prediction tasks. 
They show that across 16 global equity markets and 13 languages, LLM-based representations generate superior performance in forecasting returns. 
Their findings underscore that rich text representations, beyond surface counts or sentiment scores, can capture incremental predictive information.

Nevertheless, LLM embeddings typically operate at the document level, and do not explicitly connect content across languages. 
In multilingual news corpora, matching semantically similar passages remains a challenge: naive cosine similarity-based methods tend to be dense and uninformative, and statistical alignments may fail when the coverage diverges across languages.

In this paper, we propose a sentence-level alignment method based on optimal transport, enabling more interpretable and sparse maps between languages. 
Concretely, we embed sentences from English and foreign language articles using a multilingual encoder (LaBSE \cite{labse}), then compute an optimal transport plan to align semantically close sentences. 
The resulting sparse alignment highlights which sentence pairs truly share similar meanings.
We then aggregate aligned, unaligned, and full-text embeddings to build return scores and test them in long-short strategies. 
The long-short portfolios based on aligned embeddings yield an idealized Sharpe ratios of 4.36, outperforming the embeddings based on full or unaligned texts.

\section{Related Work}

The most closely related study is \citet{lost-in-translation}, which uses pre-calculated sentiment scores from Thomson Reuters News Analytics for English and Japanese news to examine how language-specific sentiment predicts movements in the Japanese stock market. In contrast, we provide interpretable, sentence-level alignments and link English and Japanese news content to the cross-section of Japanese stock returns.

Statistical machine translation (SMT) explicitly aligns words or phrases to facilitate translation. Early methods perform alignments at various granularities—word, phrase, or sentence \cite{brown1988statistical, gale-church-1993, ibmmodel, gal2013ibm}—but typically assume one-to-one correspondences. In contrast, multilingual news about the same company may cover different aspects and lack direct correspondence. Our goal is not translation, but to identify semantically similar or distinct content that is informative for stock return prediction. Ideas from SMT, however, motivate our approach to sentence-level alignment.

Recently, \citet{ot-alignment} use optimal transport to align words across languages for improved translation and cross-lingual transfer learning, fine-tuning models on short parallel sentences. We extend this approach to sentence-level alignment in longer documents. Longer articles often contain many stopwords, which can dilute word-level embeddings, whereas sentence-level alignment better captures semantic meaning. A major challenge is the lack of ``gold standard'' alignments at scale, which we address by leveraging optimal transport for unsupervised, interpretable sentence alignment applicable to financial prediction.

\section{Data}
We collect global news articles from the Bloomberg terminal news feed covering the period 2008–2024.\footnote{Bloomberg provided us access to the data under a non-disclosure agreement.} 
For each unique story identifier, we retain only the final update published within 24 hours of its initial appearance on Bloomberg. 
We then align news articles with stock prices and returns based on release times: articles published between 30 minutes before market open on day $t-1$ and 30 minutes before market open on day $t$ are associated with day $t$'s prices and returns. For example, the Tokyo market opens at 9:00 am, an article released at 8:28 am on day $t$ is linked to the stock price and return on day $t$, while an article released at 8:50 am on day $t$ is associated with the stock price and return on day $t+1$.
In our experiments, we include English and foreign-language articles that are explicitly associated with a single stock ticker with a ticker score $\geq 75$, indicating medium-to-high relevance to that stock.
For each stock $s$ and trading day $t$, we concatenate all associated English articles into a single composite article $\calE_{t,s}$, and all associated foreign-language articles into $\calF_{t,s}$. 
We focus primarily on the Japanese market, which has the richest foreign-language coverage in the news feed. In the preprocessed dataset, this corresponds on average to approximately 35,000 English articles and 27,000 Japanese articles annually, spanning about 3,500 stocks traded on the Tokyo Stock Exchange. 
By comparison, the Hong Kong, Taiwan, and mainland Chinese exchanges yield fewer than 10,000 English articles and 3,700 Chinese (simplified or traditional) articles annually, covering fewer than 1,000 stocks in each exchange, starting from 2012, while the Chinese coverage ranks the second among non-English news in the dataset. 
After concatenation in the Japanese market, we obtain an average of 17,807 stock-days\footnote{A stock-day is defined as a trading day for a given stock.} per year with both English and Japanese news available. 
Stock return data are obtained from Compustat Global using WRDS for the period 2008–2024.

\section{Methodology}

Our objective is to align English and foreign-language articles at the sentence level for each stock-day. We seek to identify pairs of sentences that capture similar semantic content, while minimizing misalignments, thereby enhancing interpretability of the results.

\paragraph{Text Preprocessing} 
Starting from the raw dataset, we retain only articles whose body text is updated within 24 hours of their initial publication on Bloomberg. For each article, we remove Bloomberg-specific headers and footers, keeping only the main body. Within each paragraph, line breaks are removed to ensure continuous text. We do not remove numbers or stopwords, since they may help preserve sentence structure when processed by machine learning models. Articles shorter than 100 characters or longer than 100,000 characters in any language are excluded.

\paragraph{Sentencizing and Embedding} 
We use spaCy \cite{spacy} monolingual models to split each article into sentences, yielding \\
$\calE_{t,s}=\bs{E_{t,s,1},...,E_{t,s,n}}$ for English articles and $\calF_{t,s}=\bs{F_{t,s,1},...,F_{t,s,m}}$ for foreign-language articles. 
Each sentence is then embedded using the pre-trained LaBSE model \cite{labse}, a language-agnostic encoder that assigns similar embeddings to semantically equivalent text across languages. The resulting embeddings are normalized and stacked into matrices $X_{t,s}^{E} \in \RR^{n\times 768}$ and $X_{t,s}^{F} \in \RR^{m\times 768}$ for English and foreign language articles, respectively. 

\paragraph{Optimal Transport for Alignments} 
Optimal transport (OT) \cite{monge1781, kantorovich1942, Santambrogio2015} provides a principled way to map probability mass from one distribution to another while minimizing transport cost. The original Monge formulation \cite{monge1781} is often intractable and may not even admit a solution unless restrictive conditions are met \cite{cfm2002}. Instead, we adopt the Kantorovich formulation \cite{kantorovich1942}, and in particular its discrete version as described in \cite{cuturi2020, ot-alignment}. Let $\bs{x_i}_{i=1}^n$ and $\bs{y_j}_{j=1}^m$ denote two point sets with associated probability distributions $p_x$ and $p_y$, where $\sum_{i=1}^n p_{x_i}=1$ and $\sum_{j=1}^m p_{y_j}=1$. The cost of moving mass from $x_i$ to $y_j$ is given by $c_{ij}=c(x_i, y_j)$. The optimal transport problem seeks a transport plan $\gamma_{ij}\geq 0$ that minimizes the total cost of moving probability mass:
\[\min_{\gamma} \bs{\sum_{ij} c_{ij} \gamma_{ij} : \gamma_{ij}\geq 0, \sum_i \gamma_{ij} = p_{y_j}, \sum_j \gamma_{ij} = p_{x_i}}\]
The resulting transport plan is self-normalized and sparse \cite{ot-alignment,swanson2020}, making it more effective than pure cosine similarities. In practice, the optimal $\gamma$ is approximated via entropic regularization using the Sinkhorn algorithm~\cite{sinkhorn}. Computational efficiency can be further improved by exploiting sparse and low-rank matrix structures \cite{sinkhorn-splr}, yielding up to a tenfold speedup in our setting and making large-scale analysis feasible. 
A large value of $\gamma_{ij}$ indicates that $x_i$ and $y_j$ are likely to have similar semantics.

In this paper, we follow \citet{ot-alignment} but extend the method from word-level to sentence-level alignments. Specifically, the English and foreign-language articles, $X_{t,s}^E$ and $X_{t,s}^F$, are treated as two high-dimensional point sets consisting of $n$ and $m$ each. Sentences are assigned equal probability mass, \textit{i.e.}, uniformly distributed. The cost matrix is defined using pairwise cosine distance,
\[c_{ij} = 1 - X_{t,s,i}^E \cdot X_{t,s,j}^F\]
and scaled to $[0,1]$ with min-max normalization. 

We compute transport plans in both directions: source-to-target (foreign to English) and target-to-source (English to foreign), denoted $\gamma$ and $\gamma'$. 
For each row $i$ in $\gamma$, we identify the maximum element $\gamma_{ij_{max}}$. If $\gamma_{ij_{max}}$ falls within the top 5\% of column $j_{max}$, we mark a forward alignment, producing the binary forward alignment matrix $A\in \bs{0,1}^{n\times m}$. The backward alignment matrix $B\in \bs{0,1}^{m\times n}$ is computed analogously.\footnote{Due to sparsity, one could instead apply a global threshold to directly enforce one-to-one mappings.} 
Because news coverage varies across languages, it is possible that for a given stock-day, content reported in one language is not covered in the other. In such cases, OT may still produce spurious alignments. To filter these out, we use the raw pairwise cosine similarity matrix $\xi_{ij} = X_{t,s,i}^E \cdot X_{t,s,j}^F$, and retain only alignments where $\xi_{ij} \geq \xi_{thres}$, where $\xi_{thres}$ is a hyperparameter that is dependent on the distribution of cosine similarities of the embedding model. We choose $\xi_{thres}=0.6$ for LaBSE. 
The final alignment matrix is obtained as the intersection of three matrices:
\[\calA = A * B^T * (\xi_{ij}\geq \xi_{thres}),\]
where $*$ denotes elementwise multiplication.

\paragraph{Aggregating Embeddings} 
Once the alignment matrix $\calA$ is obtained, we aggregate sentence embeddings based on alignment status. 
Specifically, we average the embeddings of aligned sentences ($\calA_{ij}=1$) in each language to obtain the aligned embeddings $X_{t,s}^{E,A}$ and $X_{t,s}^{F,A}$, and similarly average unaligned senteces ($\calA_{ij}=0$) to obtain $X_{t,s}^{E,UA}$ and $X_{t,s}^{F,UA}$. We also compute the average embeddings over all sentences to produce global embeddings $X_{t,s}^{E,Full}$ and $X_{t,s}^{F,Full}$. Each aggregated embedding is a 768-dimensional vector.
For notation, we denote these aggregated embeddings by $X_{t,s}^{l, k}$, where $l\in \bs{E, F}$ indicates the language and $k\in\bs{A, UA, Full}$ indicates the aggregate type. The superscript Full may be omitted.

\paragraph{Return Score Constructions} 
Let $\text{Ret}_{t,s}^{OC}$ denote the open-close return for stock $s$ on day $t$. We use Ridge regression with weights $w$ to associate return scores to the embeddings, following \cite{ckx}:
\[\arg\min_w \norm{X_{t,s}^{l,k}w - \text{Ret}_{t,s}^{OC}}_2^2 + \lambda \norm{w}_2^2,\]
with regularization parameter $\lambda$ selected by cross validation over the search space $[10,20,30,...,100]$, with optimal values typically  chosen as 90 or 100.
We adopt a rolling window framework: for each year $y$, the model is trained on data from year $y-5$ through $y$ and then used to generate return scores $\hat{\text{Ret}}_{t,s}^{OC,l,k}=X_{t,s}^{l,k}w$ (denoted $\text{Soft}_{t,s}^{l,k}$) for day $t$ in year $y+1$. This procedure, repeated for years 2012–2024, ensures that return scores are always computed using past data, avoiding look-ahead bias. The analysis period for evaluation is 2018–2024.

\section{Results}

\begin{figure*}[!htb]
  \centering
  \begin{subfigure}[b]{0.32\linewidth}
  \centering
  \includegraphics[width=\linewidth]{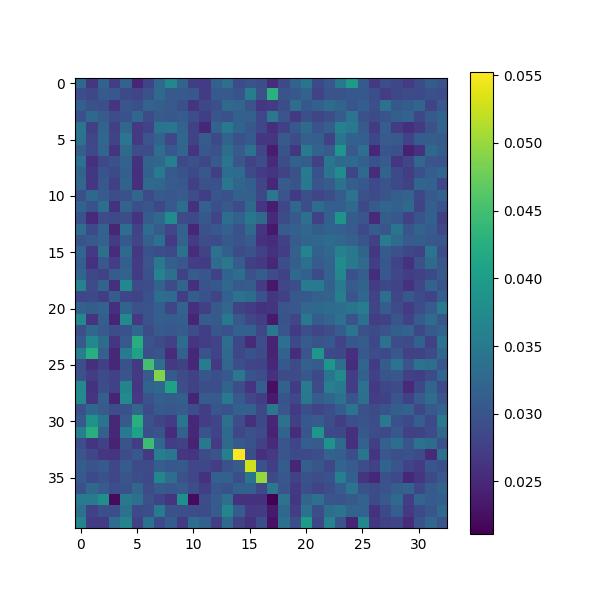}
  \caption{Softmax (Cosine Similarity)}
  \end{subfigure}
  \hfill
  \begin{subfigure}[b]{0.32\linewidth}
  \centering
  \includegraphics[width=\linewidth]{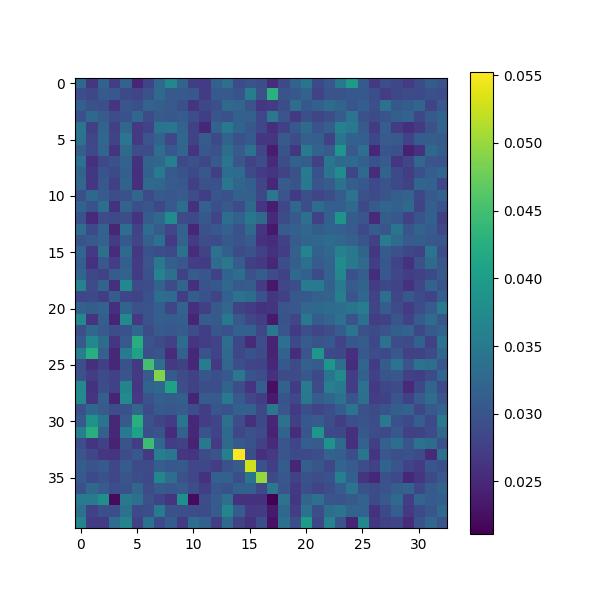}
  \caption{Entmax (Cosine Similarity)}
  \end{subfigure}
  \hfill
  \begin{subfigure}[b]{0.32\linewidth}
  \centering
  \includegraphics[width=\linewidth]{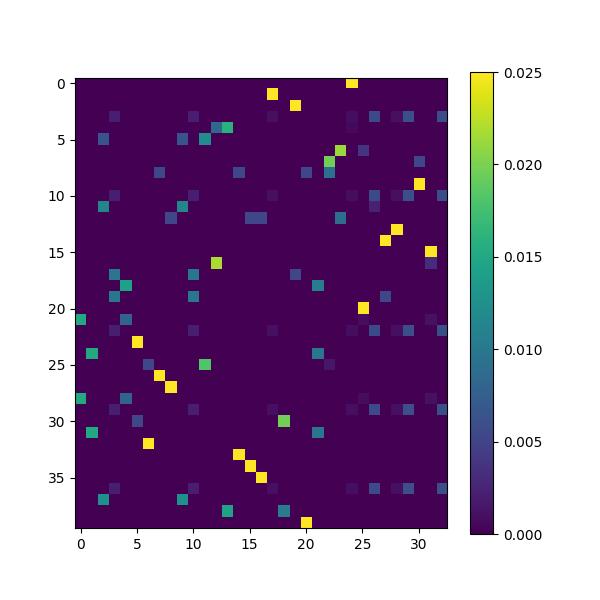}
  \caption{Optimal Transport}
  \end{subfigure}
  \vspace{-0.3cm}
  \caption{Sparsity of Different Alignment Methods}\label{fig:alignment-sparsity}
  \Description{Sparsity of Different Alignment Methods}
\end{figure*}

\paragraph{Alignment Sparsity} 
Figure~\ref{fig:alignment-sparsity} illustrates the sparsity of different alignment methods applied to news articles discussing the Bank of Japan (8301.T) on 2023-01-04. We compare pure cosine similarities, normalized via Softmax or Entmax \cite{entmax}, and optimal transport. Both Softmax and Entmax produce dense  transport plan matrices, whereas optimal transport yields a sparse matrix with non-zero values concentrated in relatively few locations. This sparsity facilitates the selection of a global threshold for filtering aligned sentence pairs. Some examples of aligned sentences are provided in the Appendix. In the dataset, updates to a single news article may be reported as several stories with slight rephrasing, resulting in multiple possible alignments.

\begin{figure*}[!htb]
  \centering
  \begin{subfigure}[b]{0.48\linewidth}
  \centering
  \includegraphics[width=\linewidth]{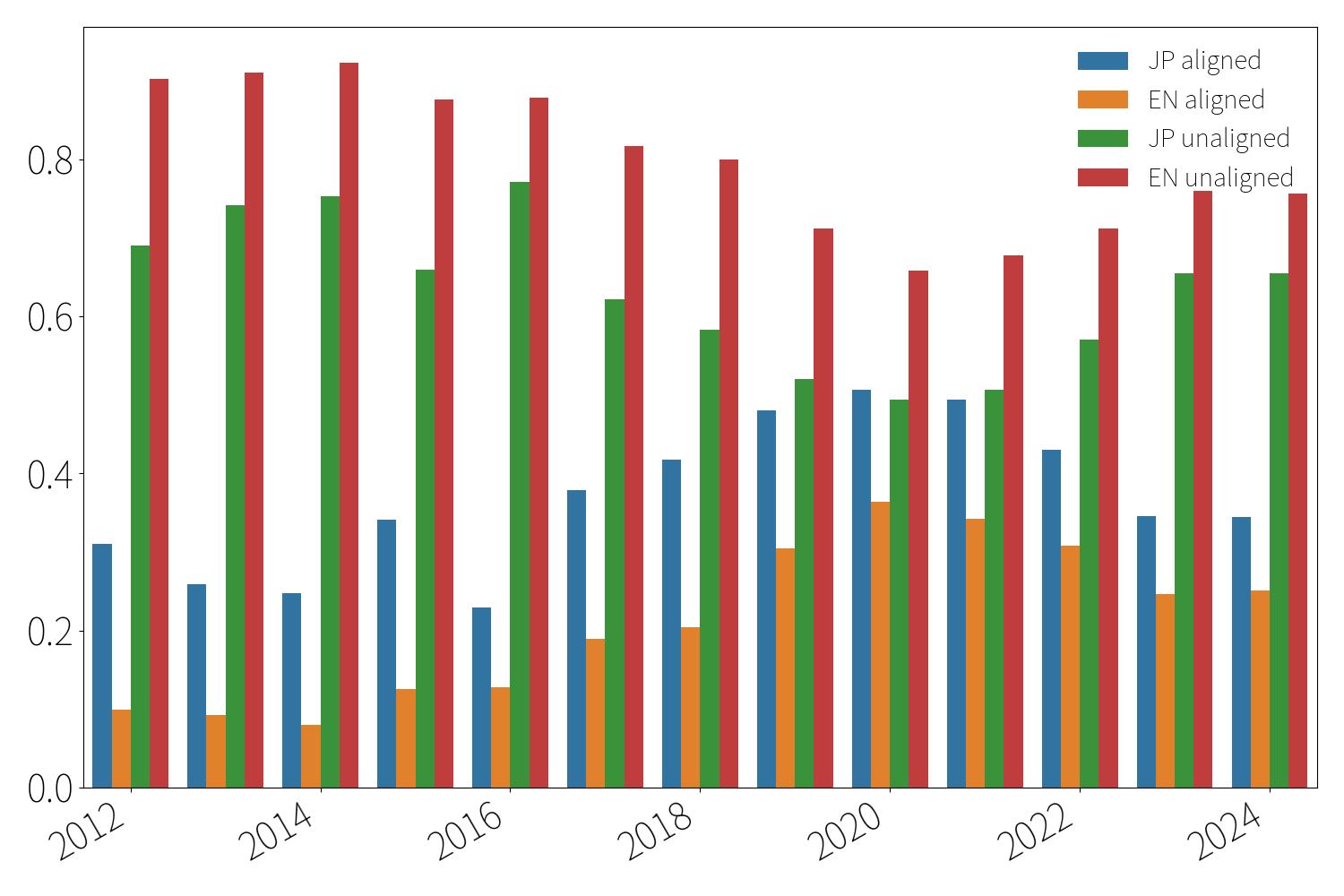}
  \caption{Proportion of Sentences Aligned}
  \end{subfigure}
  \hfill
  \begin{subfigure}[b]{0.48\linewidth}
  \centering
  \includegraphics[width=\linewidth]{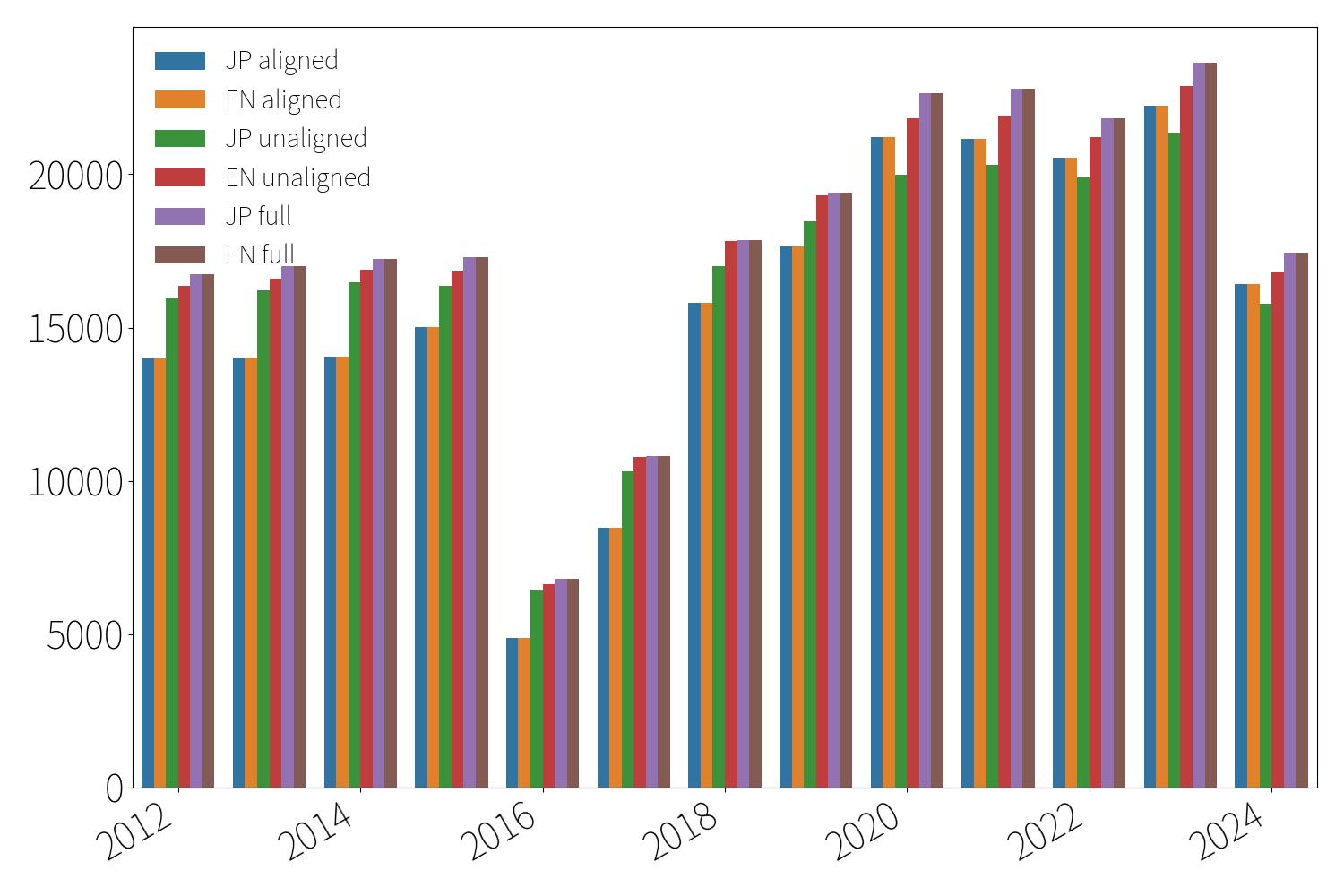}
  \caption{Number of Stock-days with Aligned and Unaligned Sentences}
  \end{subfigure}
  \vspace{-0.3cm}
  \caption{Alignments Found for Japanese News Articles}\label{fig:alignment-found}
  \Description{Proportion of sentences aligned and unaligned.}
\end{figure*}

Figure~\ref{fig:alignment-found}(a) reports the average proportion of aligned and unaligned sentences within each article, where ``JP'' refers to Japanese news and ``EN'' to English news. Overall, approximately 30–50\% of sentences are aligned in more recent years. Note that spaCy’s sentencizing may occasionally be inaccurate, particularly for articles containing bullet points, which can affect the total sentence count. 
Figure~\ref{fig:alignment-found}(b) shows the number of stock-days with aligned and unaligned sentences. Most stock-days contain at least one aligned and one unaligned sentence pair.

\paragraph{Similarity of Semantics and Return Scores} 
Table~\ref{tab:cosine-similarity} summarizes the distribution of cosine similarities for embeddings of aligned ($X_{t,s}^{E,A}$ and $X_{t,s}^{F,A}$), unaligned ($X_{t,s}^{E,UA}$ and $X_{t,s}^{F,UA}$) and full articles ($X_{t,s}^{E}$ and $X_{t,s}^{F}$) across the entire data sample from 2012 to 2024. 
As one would expect: 
(1) Aligned embeddings exhibit the highest average cosine similarity $0.8$, with low variance, reflecting strong semantic similarity.
(2) Unaligned embeddings generally show lower cosine similarities, averaging around 0.53 with higher standard deviation. Potentially, the current alignment parameters may be too strict, producing false negatives.
(3) Full-article embeddings fall in between, with an average similarity of $0.75$.
Varying the cosine similarity cutoff thresholds has minimal impact, indicating that the optimal transport-based alignment robustly identifies semantically similar content.

Table~\ref{tab:return-score-correlation} reports correlations between return scores computed from different aggregated embeddings. Given that we work with daily returns across more than 70,000 observations from 2018 to 2024, it is challenging to achieve high correlations with realized returns. 
Nevertheless, consistent patterns emerge: return scores derived from aligned embeddings ($\text{Soft}^{l,A}$) tend to exhibit higher correlations, while scores from unaligned embeddings ($\text{Soft}^{l,UA}$) show lower correlations.

\begin{table}[!htb]
\caption{Sentence Level Cosine Similarity}\label{tab:cosine-similarity}

\centering
\resizebox*{0.75\linewidth}{!}{\input{tables/JP/tables_0.95/sentence_cosine_sim.tex}}
\end{table}

\begin{table}[!htb]
\caption{Return Score Correlations}\label{tab:return-score-correlation}

\centering

\resizebox*{\linewidth}{!}{\input{tables/JP/tables_0.95/sentence_LaBSE_ret_one_day_oc_0_corr.tex}}
\end{table}

\paragraph{Impact on Trading Strategy} 
To assess whether commonly discussed information influences investor behavior and market performance, we implement a long-short trading strategy based on the constructed return scores. On each trading day with at least 20 traded stocks, we rank the stocks by their predicted return scores $\text{Soft}_{t,s}^{l,k}$, and divide them into quantiles. Long the stocks in the top-quantile and short the stocks in the bottom-quantile. Let $\text{Ret}_t^{OC, L,l,k}$ and $\text{Ret}_t^{OC, S,l,k}$ denote the average long and short returns on day $t$, for each language $l$ and alignment type $k$. The long-short return is defined as
\[\text{LS}_t^{l,k} = \text{Ret}_t^{OC, L,l,k} - \text{Ret}_t^{OC,S,l,k}.\]
We compute the distributional statistics of $\text{LS}_t^{l,k}$, as well as Sharpe ratios for the resulting portfolios. 
Table~\ref{tab:strategy-summary} summarizes the results. 
The geometric mean of the daily long-short returns is calculated as
\[\bb{\prod_{t=1}^T (1+\text{LS}_t^{l,k})}^{1/T}-1,\] 
where $T$ is the total number of trading days from 2018 to 2024. The daily Sharpe ratio is computed as $\frac{\text{mean}(\text{LS}_t^{l,k})}{\text{std}(\text{LS}_t^{l,k})}$, and the annualized Sharpe ratio is daily Sharpe ratio multiplied by $\sqrt{252}$. 
In the Japanese market, the full embedding of Japanese news generates higher Sharpe than English news. This means that Japanese news typically have higher association with the returns in Tokyo exchange. 
In the Japanese market, portfolios based on the full embeddings of Japanese news achieve higher Sharpe ratios than those based on English news, indicating that Japanese news generally has stronger predictive power for returns on the Tokyo Stock Exchange. Portfolios constructed from aligned embeddings exhibit even higher Sharpe ratios for both Japanese and English texts, suggesting that sentences capturing common themes across languages provide clearer signals of stock performance. In contrast, unaligned embeddings tend to be noisier and less informative.

\begin{table}[!htb]
\caption{Strategy Summary}\label{tab:strategy-summary}

\centering

\resizebox*{\linewidth}{!}{\input{tables/JP/tables_0.95/strategy_stats_T_0_daytime_returns_LaBSE.tex}}
\end{table}

\section{Conclusion}

In this paper, we use optimal transport to align semantically similar sentences in multilingual news articles. Compared to pure cosine similarities, optimal transport produces sparser and more interpretable alignments. Sentences identified as aligned exhibit high semantic similarity, and return scores derived from these aligned sentences show stronger correlations with realized returns. In a long-short trading strategy, portfolios based on aligned sentences achieve higher Sharpe ratios, indicating that commonly discussed themes across languages provide more informative signals for predicting stock returns.
Future directions for this research include: (1) extending the approach to additional markets and languages, and (2) improving thresholding techniques to reduce false negatives in the alignment process.

\begin{acks}
This research was carried out at Rotman School of Management, FinHub Financial Innovation Lab, University of Toronto. We gratefully acknowledges financial support and insightful discussions from Royal Bank of Canada (RBC) Capital Markets. New York, NY and Toronto, ON, Canada.
\end{acks}

\bibliographystyle{ACM-Reference-Format}
\bibliography{main}

\appendix
\section{Sample Aligned Sentences}

In this appendix, we compare the sentence alignments produced by Softmax, Entmax, and Optimal Transport for a Japanese news article about the Bank of Japan (8301.T) published on 2023-01-04. For Softmax and Entmax, we select alignments corresponding to the top 5\% probabilities across the entire matrix. The results are summarized in Table~\ref{tab:sample-aligned-sentences}, where correct alignments are manually identified and highlighted in bold. We also observe that semantic similarity plays the key role: aligned pairs may differ in numerical values, or in some cases, involve sentences that are not direct translations but still convey related meanings.

\begin{table*}[!htb]
\caption{Sample Aligned Sentences}\label{tab:sample-aligned-sentences}

\centering
\includegraphics[width=\linewidth]{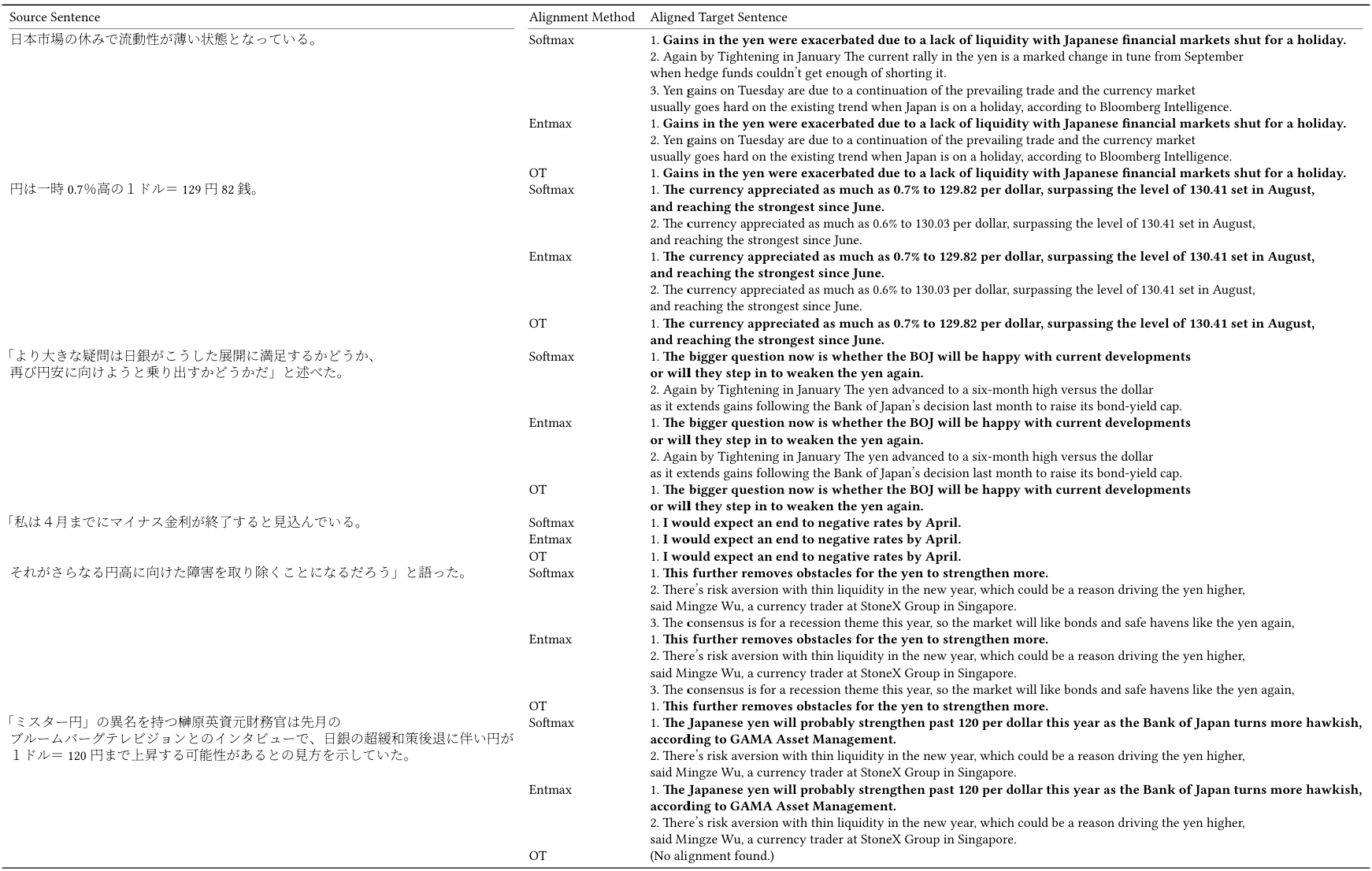}
\end{table*}

\end{document}

%% file: tables/JP/tables_0.95/sentence_cosine_sim.tex
\begin{tabular}{lccccc}
\toprule
Alignment & mean & std & 5\% & 50\% & 95\% \\
\cmidrule(lr){2-6}
Aligned & 0.79 & 0.06 & 0.66 & 0.81 & 0.87 \\
Unaligned & 0.53 & 0.20 & 0.18 & 0.54 & 0.81 \\
Full & 0.75 & 0.09 & 0.59 & 0.76 & 0.86 \\
\bottomrule
\end{tabular}

%% file: tables/JP/tables_0.95/sentence_LaBSE_ret_one_day_oc_0_corr.tex
\begin{tabular}{lccccccc}
\toprule
 & Ret & Soft$^{EN, A}$ & Soft$^{EN, UA}$ & Soft$^{EN}$ & Soft$^{JP, A}$ & Soft$^{JP, UA}$ & Soft$^{JP}$ \\
\midrule
Ret & 1.00 & 0.02 & 0.03 & 0.03 & 0.02 & 0.01 & 0.03 \\
Soft$^{EN, A}$ & 0.02 & 1.00 & 0.41 & 0.67 & 0.67 & 0.38 & 0.57 \\
Soft$^{EN, UA}$ & 0.03 & 0.41 & 1.00 & 0.76 & 0.49 & 0.47 & 0.50 \\
Soft$^{EN}$ & 0.03 & 0.67 & 0.76 & 1.00 & 0.62 & 0.41 & 0.62 \\
Soft$^{JP, A}$ & 0.02 & 0.67 & 0.49 & 0.62 & 1.00 & 0.44 & 0.72 \\
Soft$^{JP, UA}$ & 0.01 & 0.38 & 0.47 & 0.41 & 0.44 & 1.00 & 0.71 \\
Soft$^{JP}$ & 0.03 & 0.57 & 0.50 & 0.62 & 0.72 & 0.71 & 1.00 \\
\bottomrule
\end{tabular}

%% file: tables/JP/tables_0.95/strategy_stats_T_0_daytime_returns_LaBSE.tex
\begin{tabular}{llrrrrrrrr}
\toprule
Alignment & Lang & Geo Mean & Mean & Std & 5\% & 50\% & 95\% & Sharpe & Ann. Sharpe \\
\cmidrule(lr){1-2} \cmidrule(lr){3-10}
Aligned & JP & 0.35\% & 0.36\% & 1.32\% & -1.76\% & 0.39\% & 2.31\% & 0.27 & 4.36 \\
 & EN & 0.28\% & 0.29\% & 1.34\% & -1.86\% & 0.23\% & 2.46\% & 0.22 & 3.42 \\
Unaligned & JP & 0.17\% & 0.18\% & 1.34\% & -1.98\% & 0.16\% & 2.29\% & 0.13 & 2.12 \\
 & EN & 0.23\% & 0.24\% & 1.29\% & -1.66\% & 0.16\% & 2.45\% & 0.18 & 2.91 \\
Full & JP & 0.30\% & 0.31\% & 1.23\% & -1.59\% & 0.27\% & 2.38\% & 0.25 & 3.98 \\
 & EN & 0.24\% & 0.25\% & 1.16\% & -1.65\% & 0.17\% & 2.20\% & 0.21 & 3.40 \\
\bottomrule
\end{tabular}